\pdfoutput=1
\documentclass[11pt,twoside]{article}
\usepackage{asp2010}
\begin{document}

\title{Addressing the Photometric Calibration Challenge: Explicit Determination of the Instrumental Response and 
Atmospheric Response Functions, and Tying it All Together.}
\author{Christopher W.~Stubbs$^1$, John L.~Tonry$^2$
\affil{$^1$Department of Physics, Harvard, 17 Oxford St. Cambridge MA 02138 USA}
\affil{$^2$Institute for Astronomy, Univ. of Hawaii, 2680 Woodlawn Drive, Honolulu, HI 96822-1839, USA}
}

\begin{abstract}

Photometric calibration is currently the dominant source of systematic uncertainty in exploiting type Ia 
supernovae to determine the nature of the dark energy.  We review our ongoing program to address this calibration challenge 
by performing measurements of both the instrumental response function and the optical transmission function of the
atmosphere. A key aspect of this approach is to complement standard star observations by using NIST-calibrated photodiodes as a metrology foundation 
for optical flux measurements. We present our first attempt to assess photometric consistency between synthetic photometry and observations, 
by comparing predictions based on 
a NIST-diode-based determination of the PanSTARRS-1 instrumental response and empirical atmospheric transmission measurements, 
with fluxes we obtained from observing spectrophotometric standards.

\end{abstract}

\section{Introduction and Motivation}

The challenge of photometry is to extract knowledge of the location and flux distribution of 
astronomical sources, based on measurements of the 2 dimensional distribution of detected photons
in a focal plane. Each pixel $i$ in the detector array sees a signal $S_i$ given by 

\begin{equation}
S_i = \sum_{sources~j} \int \Phi_j(\lambda) R_i(\lambda) T(\lambda)  G(\lambda) A_i ~ d\lambda,
\label{eq:psignal}
\end{equation}

\noindent
where the sum is taken over all sources (including the sky) that contribute to the flux in the 
pixel, $\Phi_j(\lambda)$ is the photon spectrum for source $j$, $R_i(\lambda)$ is the 
throughput of the pixel, including the transmission of the optics and the pixel's
quantum efficiency,  $T(\lambda)$ is the optical transmission of the atmosphere, $G(\lambda)$ accounts
for non-atmospheric extinction processes along the line of sight to the source (necessary to convert from 
top-of-the-atmosphere fluxes to the SED of the source), and 
$A_i$ is the effective aperture of the system for pixel $i$, essentially the wavelength-independent
part of the instrumental response.  

Photometric calibration uncertainties currently limit our ability to use type Ia supernovae to 
determine the nature of the dark energy \citep{SNLS11}.  The essential tasks are 1) to 
establish a clear relationship between the zeropoints in the various passbands used
(to avoid contamination of the Hubble diagram with systematic zeropoint errors), and 2) to 
map out the shapes of the passbands (to enable precise K corrections). The absolute 
overall zeropoint of the system is degenerate with the intrinsic brightness of SN Ia's. Precise
colors are what matter. However we do note that in splicing together a Hubble diagram from 
multiple surveys, it is important to understand the overall zeropoints of the relevant
instrumental photometric systems. 

There are a number of alternative approaches to establishing well-understood colors in a multiband 
photometric system, comprising some combination of the following:

\begin{enumerate}

\item{} Using terrestrial sources to establish a spectral scale for celestial sources. This is the 
approach taken by \cite{Hayes75a} and \cite{Hayes75b}, that in effect uses Vega as a transfer
standard.  In this instance the radiometric properties of the terrestrial source
is used as the fundamental system calibration.  

\item{} Exploiting the theoretical understanding of stars to predict the SEDs of individual
stars. In practice white dwarfs are the favored class of object. Observations of appropriate sources are
then used to calibrate instruments, and the theoretical spectrum is the basis for the relative system calibration. 
This approach has been used for the HST calibrations described by \cite{AbsCal94,Bohlin96} and 
presented at this meeting by Hunt. This approach is also being pursued by the SkyMapper project
(as described by Bessell at this meeting). The status of theoretical spectral modeling is described by Rauch in his contribution to this meeting. 

\item{} Statistical properties of stars can also be used to establish consistency across the sky, and to correct for 
various types of chromatic attenuation. This approach was pursued by \cite{SLR}, and was used by \cite{dust11} 
to show that the commonly-used SFD \citep{SFD} extinction map ($G(\lambda)$ of equation 1) requires renormalization of 0.86 in $E(B-V)$ 
compared to SFD. 

\item{} An alternative approach, described in \cite{ST06} and in the contribution by Cramer to this meeting,
attempts to bring modern metrology methods to bear on the calibration challenge. This approach
uses NIST-calibrated detectors \citep{NIST07} as the fundamental reference for establishing the system's sensitivity function. 
Initial results from this approach were obtained on the CTIO 4 meter \citep{CTIOlasercal}, and more recently on the PanSTARRS-1
system \citep{PSlasercal}. The experimental challenge in this approach is to 
obtain a reliable measurement of the instrumental response function that is traceable to SI standards.  
 
\end{enumerate}

We stress that these approaches can all be pursued in parallel, and the relative consistency we achieve can be used
to understand, quantify, and overcome sources of systematic error. 

\section{Throughput Measurements of the PanSTARRS-1 Survey System}

\subsection{Full-aperture system response function measurements} 

\begin{figure}[!ht]
\begin{center}
\plottwo{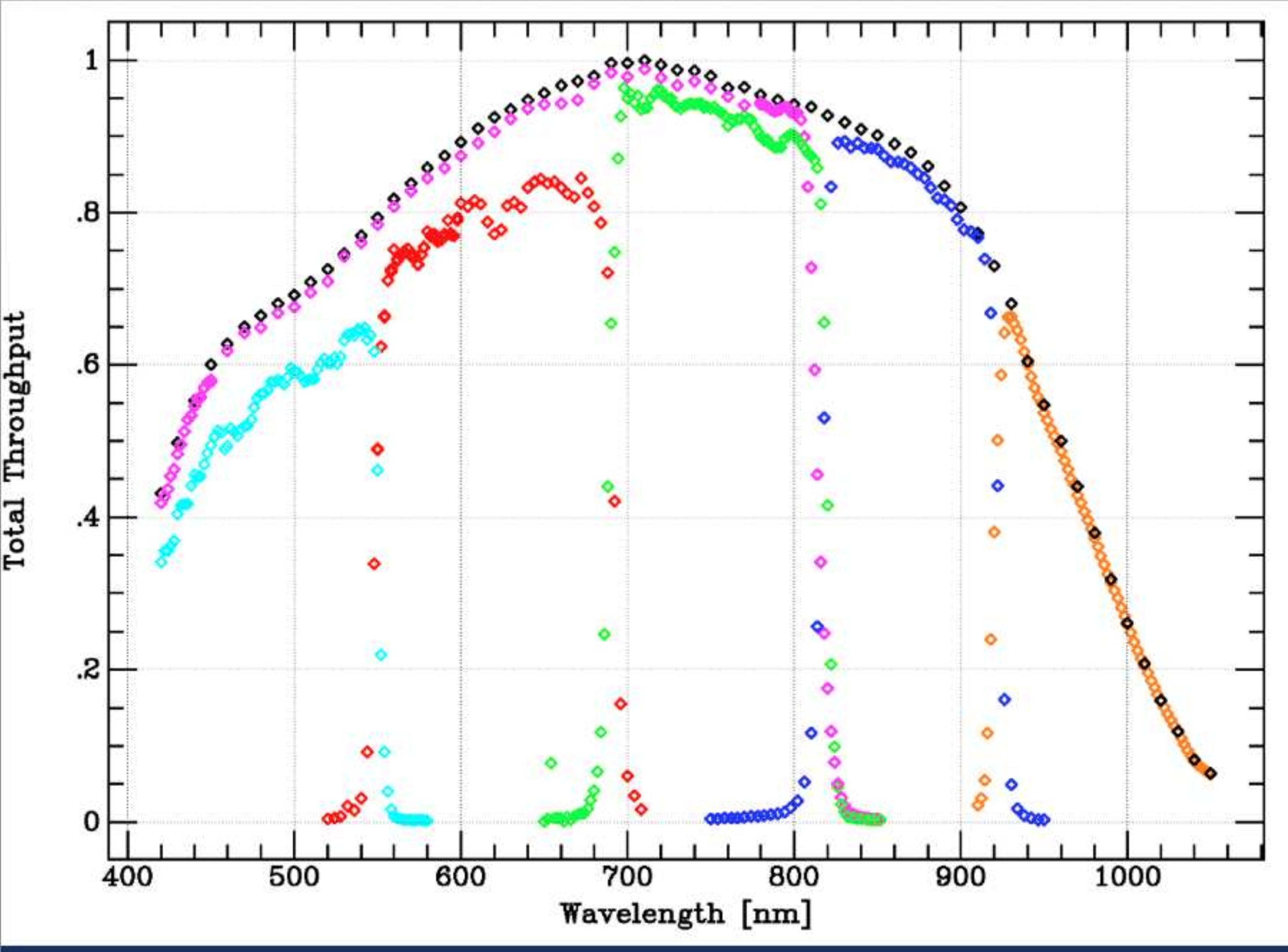}{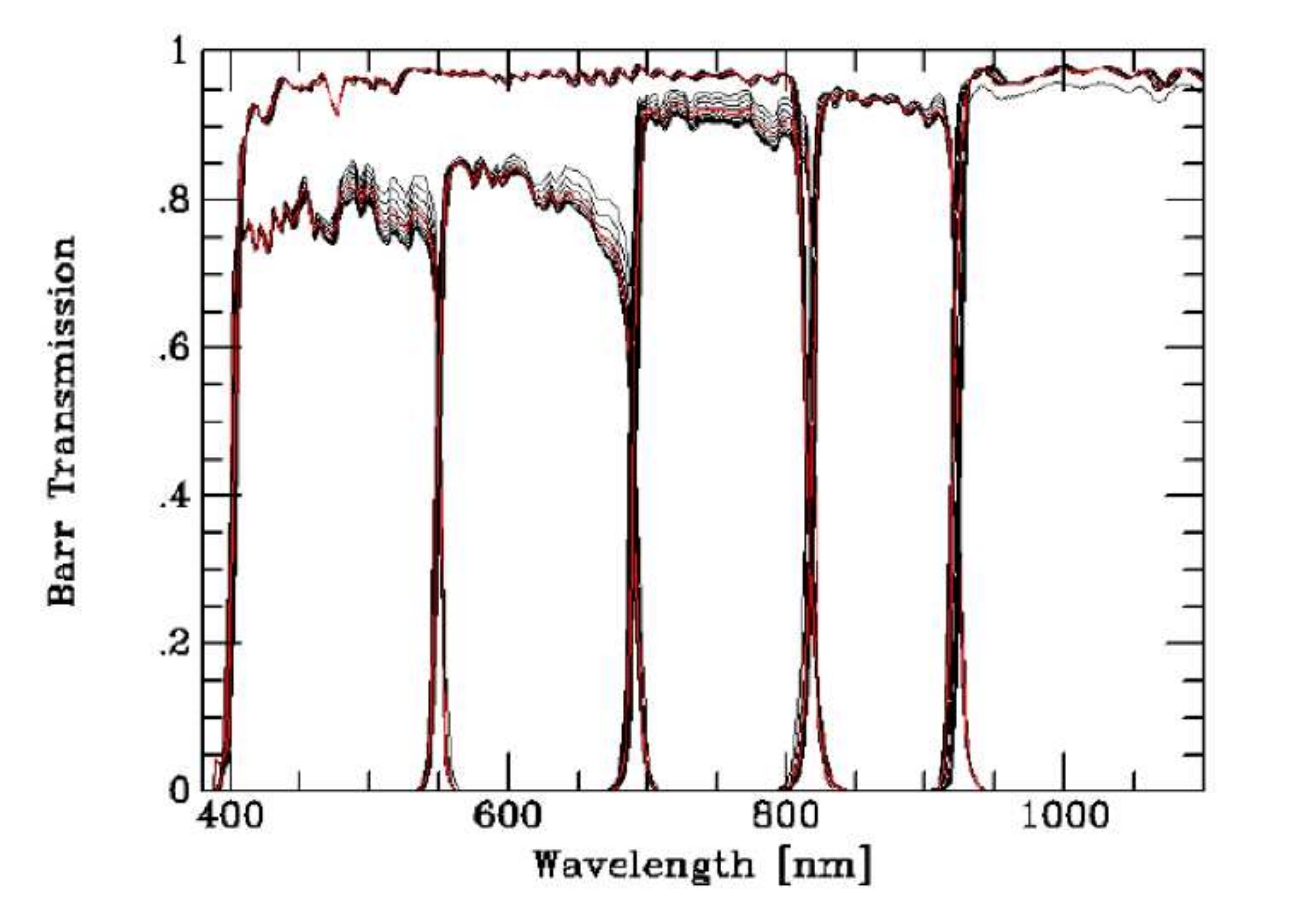}
\caption{Shown on the left is the measured relative system response function for PS-1, all normalized to a NIST photodiode. 
The panel on the right shows the radial position-dependence of the filter transmission function. Each curve corresponds to a 
different radius on the focal plane.}
\label{fig:filters}
\end{center}
\end{figure}

We have used a photodiode to monitor the flux emanating from a back-illuminated flat-field
screen in the dome of the PanSTARRS-1 telescope to map out the full-aperture relative system response function.
Details of this measurement are given in \cite{PSlasercal} and \cite{JT12}. The results are shown in Figure \ref{fig:filters}.  
We project light from a tunable laser onto the flat-field screen in the dome. We measure the 
flux emanating from the screen, incident on the telescope pupil, with a calibrated photodiode. We
then compare the flux detected by the instrument to the incident flux, as measured by the 
photodiode. Performing this measurement at a succession of wavelengths allows us to determine system 
throughput as a function of wavelength, using the calibrated photodiode as the 
fundamental reference. A philosophically similar full-aperture calibration system for the Dark Energy Survey is described in these proceedings by Marshall.  

\subsection{Collimated-Light Determination of Ghosting in the PanSTARRS-1 System}

During the processing of the throughput data described above, we became increasingly concerned about stray light, multipath effects, and ghosting
in the optical train. The basic problem is that by imaging a uniform surface brightness screen we can't distinguish the focussing light paths
through the system (which is how we measure celestial sources) from other light paths. This is the underlying reason why dome flats, twilight 
flats and sky flats typically mutually disagree, and require an ``illumination correction''. Especially at wavelengths where the filter transmission is low, 
we find a substantial amount of light scatters from the focal plane, up to the filter, and back down to the CCDs. This is a source of systematic error. The 
contribution by Regnault to this meeting describes the merits of ray tracing to model these ghosts, and the LSST calibration team is 
undertaking a similar exercise (see paper here by Lynne Jones). 

\begin{figure}[!ht]
\begin{center}
\plotone{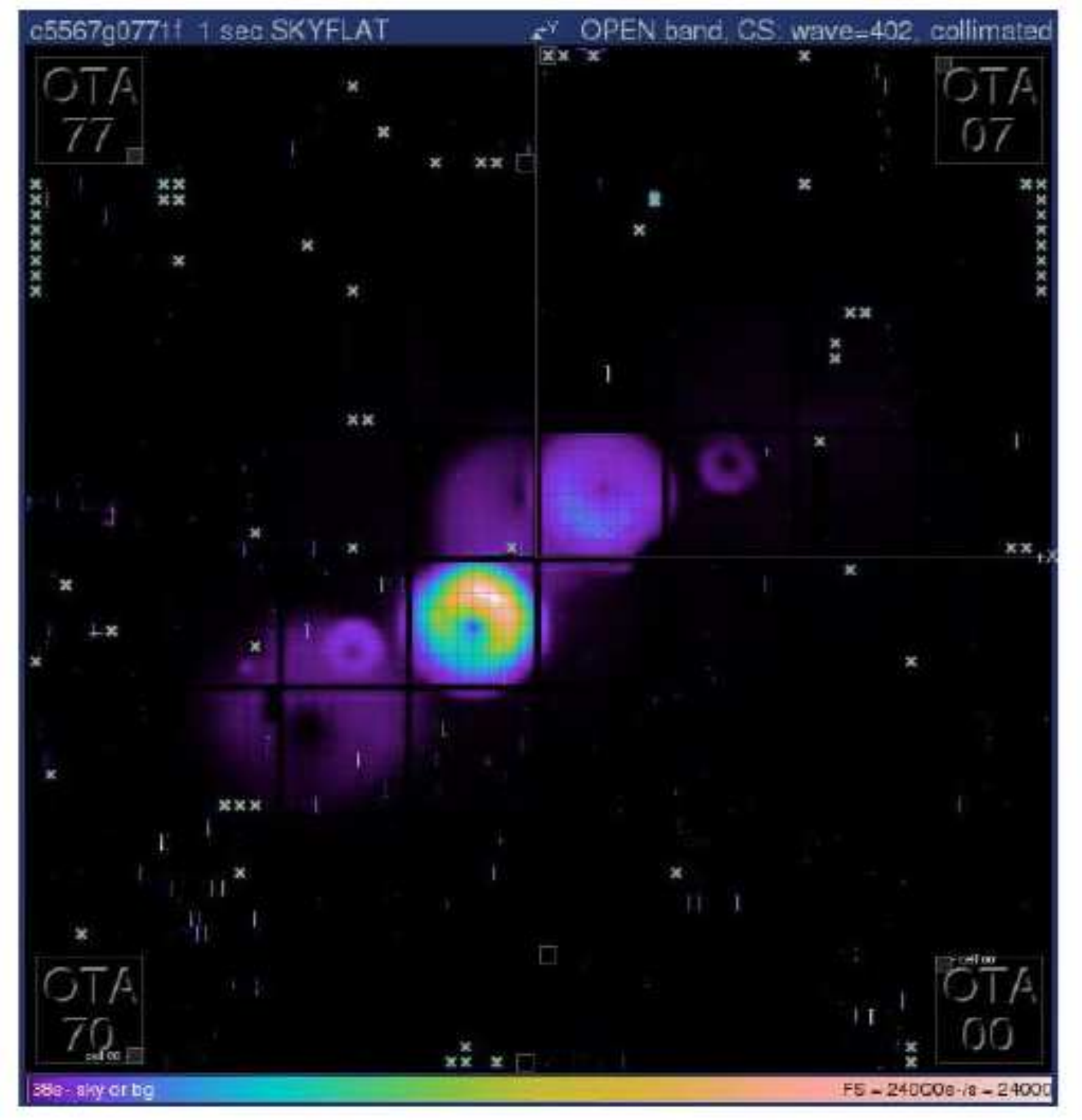}
\caption{This is a false-color image of the light distribution across the entire PanSTARRS-1 focal plane when illuminated by a
collimated beam. The lack of axisymmetry in the beam spot is because the calibration diode obscured the output pupil of the collimation telescope.
We used this determination of the scattered light vs. wavelength to obtain a more reliable measurement of relative system throughput.}
\label{fig:ghosts}
\end{center}
\end{figure}

In order to quantify the amount of ghosting in the system, we set up a collimated telescope that was fed by an optical fiber to 
send a beam of collimated light onto the primary. This controls the phase space distribution of rays entering the telescope
much better than the light emanating from the flat field screen. Moreover, we can adjust the focus of the collimating telescope to control the size
of the image on the PS-1 focal plane. Figure \ref{fig:ghosts} shows an example image obtained with this configuration.

We used the ghost-light fraction as a function of wavelength to correct the flat-screen generated throughput data. This is 
described in more detail in \cite{JT12}. 

\section{Atmospheric Transmission}

Our team's review paper on the atmosphere \citep{PASPatmos} describes the variable components of atmospheric transmission that require attention 
in order to achieve improved photometric performance. Water vapor, ozone, clouds and aerosols are the primary concerns. 
Rayleigh scattering from $O_2, N_2$ and other well-mixed gases is essentially deterministic, are the molecular absorption 
line strengths from $O_2$. Although a number of groups are conducting or are planning to undertake explicit atmospheric transmission 
monitoring (see contributions to this conference by McGraw, the Texas A\&M team, the ESO program, and Blake), 
it remains the case that our community does not yet know the angular and temporal correlation functions of these variable atmospheric attenuation processes. 
Max Fagin and Justin Albert's presentations to this conference describe our group's program to fly a set of laser diodes
to characterize aerosol extinction, so we will focus here on our measurements of water vapor. 

\subsection{Water Vapor Measurements at the \protect \hbox {Pan-STARRS1}\ site, using Polaris}

\begin{figure}[!ht]
\begin{center}
\plotone{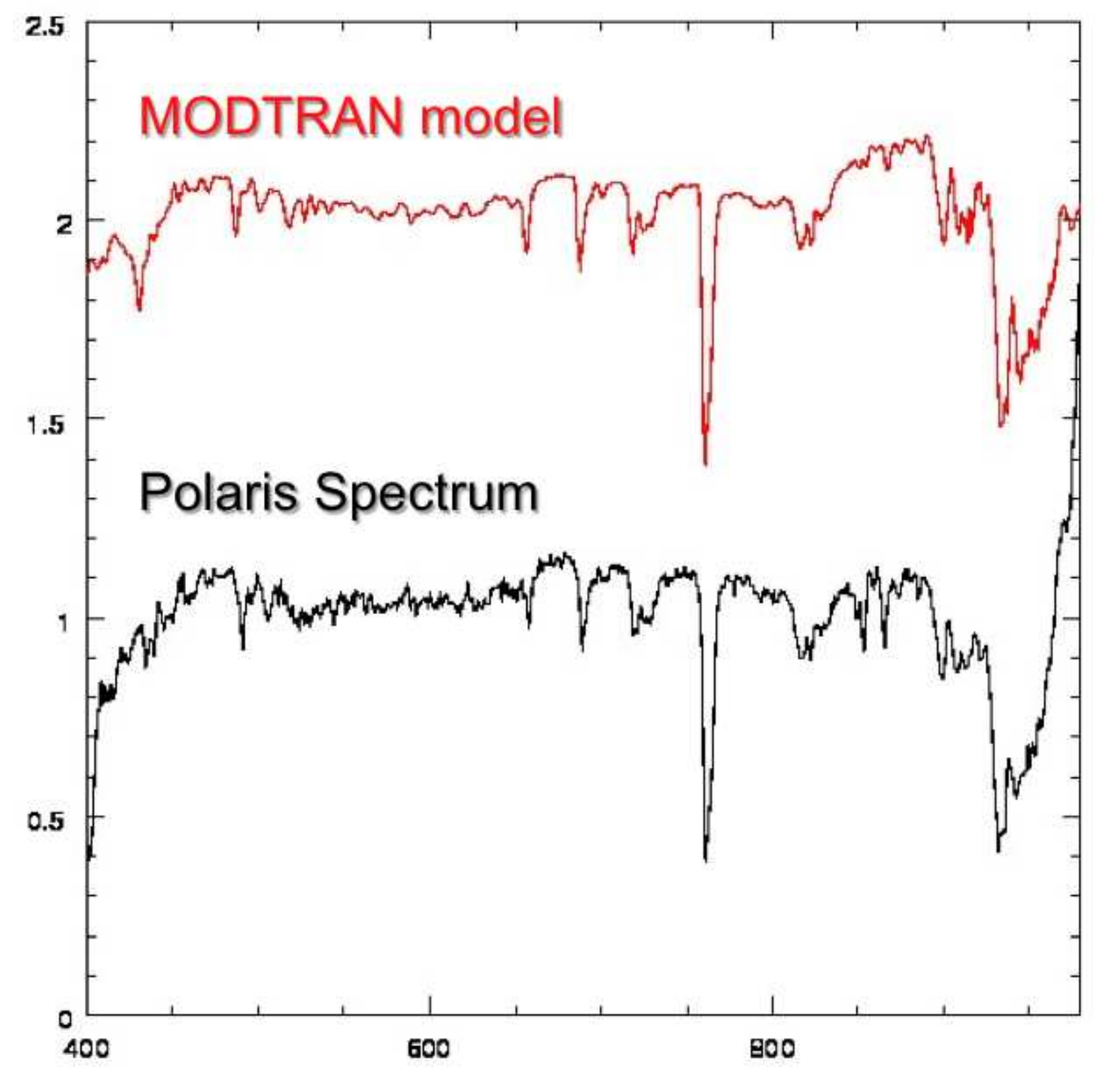}
\caption{This Figure compares the atmospheric transmission measured with the objective grating dispersed imager with 
the model atmosphere produced by MODTRAN, once we tuned the precipitable water vapor to match the equivalent widths seen in the data.
Note the Polaris spectrum also has features from the stellar atmosphere, such as the Ca triplet at 850 nm.}
\label{fig:MODTRAN}
\end{center}
\end{figure}

\begin{figure}[!ht]
\begin{center}
\plotone{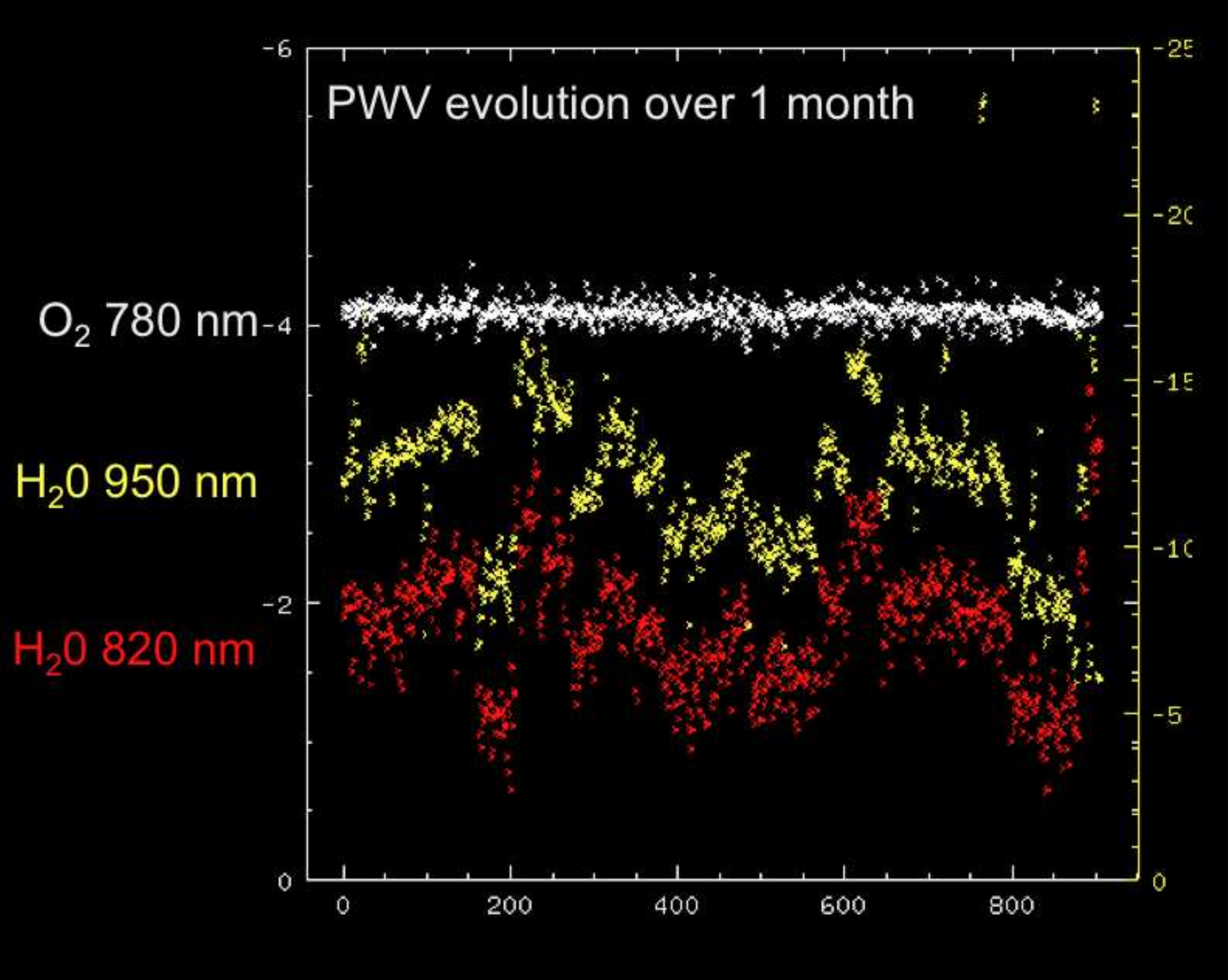}
\caption{This Figure shows the evolution of the measured equivalent widths of $O_2$ and $H_2O$ over the duration of one month. 
Both the accuracy and the precision of the oxygen line indicate that we are making a reliable measurement of atmospheric attenuation.}
\label{fig:EW}
\end{center}
\end{figure}

We have constructed an objective grating slitless imaging spectrograph (Shivvers {\it et al.~}in prep). The configuration is very similar to the
spectrophotometric system described by McGraw.  We decided to point at the North Celestial Pole, so as to obtain data at fixed airmass over an 
extended period of time. The camera in this system uses a Pixis 1024BR deep-depletion detector and so it has low fringing. 
The dispersion and focal length are such that we can obtain a spectrum spanning 300 to 1000 nm, and the 50-50 objective transmission grating 
greatly suppresses second order contamination light. Figure \ref{fig:MODTRAN} shows the comparison between one of our 1-d spectra
and the MODTRAN model that has been tuned to obtain an equivalent width of water vapor absorption to match the observations. 
The instrument was installed on Haleakala in July 2011 and has been acquiring data intermittently ever since. 
Figure \ref{fig:EW} shows the temporal evolution, over a one month period, of the equivalent widths (EW, in nm) of the $O_2$ and
water vapor features. We obtain an excellent match to the MODTRAN prediction of the EW for $O_2$, and there is no evidence
of this feature changing over the duration of the observations. Both of these facts indicate that we are obtaining meaningful and reliable 
measurements of absorption features. The water vapor attenuation does show considerable variation over this period, and we assess
our fractional accuracy in PWV determination to be about 10\%. This is adequate to obtain an overall precision of $< 1\%$ in the 
determination of transmission in the \protect \hbox {Pan-STARRS1}\ passbands.  

\subsection{MODTRAN, tweaked}

For comparison with the standard star observations, we used the MODTRAN ``Generic Tropical" model atmosphere, 
with the ``Desert Extinction (Spring-Summer)" aerosol choice. No attenuation from clouds was included.
The PWV at the base of the atmosphere was set to 0.65 cm, since this produced a good match between our Polaris
observations on the night the standard stars were observed, and the MODTRAN prediction for the line of sight from the altitude of the summit
of Haleakala, looking towards the celestial pole. As described below we ended up making an adjustment to the aerosol component, based on
the airmass dependence we observed. We note that the other, deterministic, aspects of atmospheric transmission were not adjusted. 
We used the default ozone column.  This amounted to a MODTRAN transmission model with three empirically 
adjusted parameters: PWV, aerosol optical depth, aerosol Angstrom exponent. 

\section{The Total PanSTARRS-1 System Throughput: Instrument and Atmosphere}

\begin{figure}[!ht]
\begin{center}
\plotone{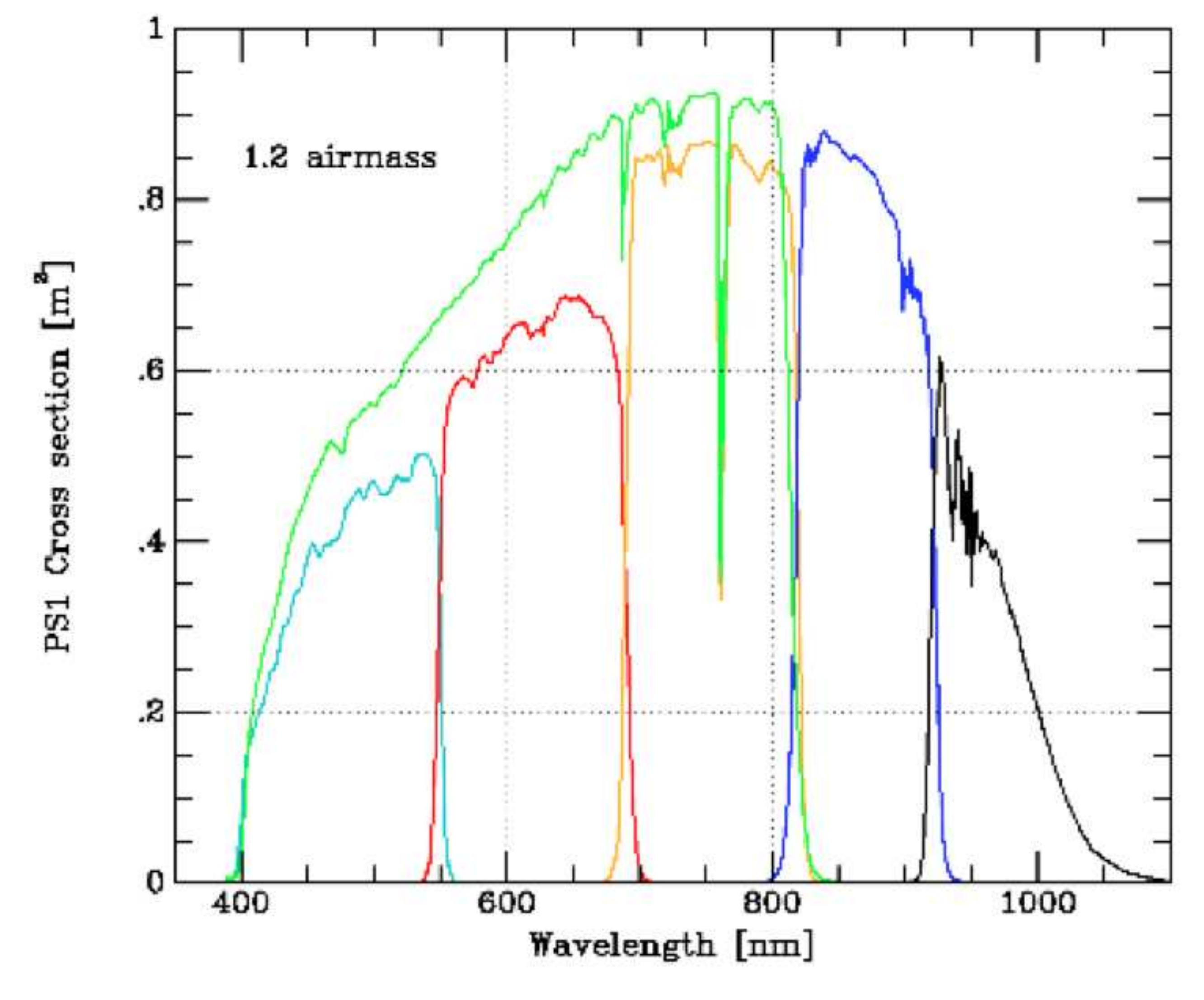}
\caption{Combining the ghost-corrected instrumental response function with the PWV-informed MODTRAN atmosphere 
at 1.2 airmasses produces the system sensitivity functions shown here.}
\label{fig:throughput}
\end{center}
\end{figure}

Taking the ghosting-corrected relative system throughput function in conjunction with the MODTRAN atmosphere (with parameters adjusted
based on our observed PWV), we obtained the PanSTARRS system throughput functions shown in Figure \ref{fig:throughput}. We have adopted
a convention of 1.2 airmasses as the definition of the PS throughput function \citep{JT12}. 

With the PS-1 system throughput function in hand, we have used measurements of HST spectrophotometric standards 
to compare our NIST-based flux calibration to the fluxes observed from these standards. Details of this are presented in 
\cite{JT12}. An important aspect of this comparison is to determine the atmospheric transmission.

For the \protect \hbox {Pan-STARRS1}\ bandpasses we integrated a set of power law SEDs against
each of these model atmospheres for each bandpass and created an
interpolation function for the extinction as a function of four
variables: $z$ for airmass ($\sec \zeta$ where $\zeta$ is the zenith
distance), $h$ for precipitable water vapor (PWV) (typically 0.65~cm at sea level),
$a$ for ``aerosol exponent'' (nominally 1; we modify the Modtran
aerosol component by applying this power to the transmission, thereby
mostly affecting the aerosol amplitude), and $p$ for SED power law (we
calculate the extinction for pure power law SEDs, where $p=+2$ for
$f_{\nu}\sim\nu^{+2}$ corresponds to an O star with $(r{-}i)=-0.43$,
$p=0$ for $f_{\nu}\sim\hbox{const}$ corresponds to an F star
with $(r{-}i)=0.00$, and $p=-2$ for $f_{\nu}\sim\nu^{-2}$ corresponds to an
K5 star with $(r{-}i)=+0.42$.  The extinction $dm$ in magnitudes 
is parameterized the extinction as  
\begin{equation}
  \ln dm = \ln C +  Z\ln z + A\ln a + P p + \ln h(H_0 + H_1\ln z + H_2\ln h)
\label{eq:modterp}
\end{equation}
The coefficients for each of the \protect \hbox {Pan-STARRS1}\ filters are given in Table~\ref{tab:modterp}.

\begin{table}[htdp]
\caption{PanSTARRS extinction coefficients. The columns contain coefficients described above for
  each of the PanSTARRS bandpasses that interpolate the Modtran extinction
  calculations.  The final column is the percentage scatter of these
  fits relative to the calculated values.  Note that the saturation of
  the water lines means that the extinction is {\it not} proportional
  to $\sec\zeta$ ($Z\neq1$), particularly for \ensuremath{y_{\rm P1}}. This is consistent with the modeling
  results from MODTRAN.}
\begin{center}
\begin{tabular}{lrrrrrrrr}
\hline
  Filter   & $C$   & $Z$   & $A$   & $P$      & $H_0$    & $H_1$    & $H_2$ & err\\
\hline                                                                                            
  \ensuremath{g_{\rm P1}}     & 0.204 & 0.982 & 0.227 & $ 0.021$ & $ 0.001$ & $-0.000$ & 0.000 & 1.7\\
  \ensuremath{r_{\rm P1}}     & 0.123 & 0.975 & 0.283 & $ 0.012$ & $ 0.012$ & $-0.000$ & 0.005 & 2.0\\
  \ensuremath{i_{\rm P1}}     & 0.092 & 0.831 & 0.304 & $ 0.005$ & $ 0.125$ & $-0.011$ & 0.035 & 2.7\\
  \ensuremath{z_{\rm P1}}     & 0.060 & 0.878 & 0.375 & $-0.004$ & $ 0.330$ & $-0.070$ & 0.055 & 4.9\\
  \ensuremath{y_{\rm P1}}     & 0.154 & 0.680 & 0.145 & $ 0.014$ & $ 0.549$ & $-0.084$ & 0.024 & 3.5\\
\hline
\end{tabular}
\end{center}
\label{tab:modterp}
\end{table}
  
\section{Standard Star Observations}

MJD 55744 (UT 02 July 2011) was a photometric night during which we observed a substantial number of spectrophotometric standard stars from the 
STIS Calspec \citep{CalSpec01} tabulation: 1740346, KF01T5, KF06T2, KF08T3, LDS749B, P177D, and WD1657- 343. These were observed 
throughout the night at airmasses between 1 and 2.2 in all six filters and also with no filter in the beam. Each observation was repeated, and 
exposure times were chosen to stay well clear of any non-linearities but still permit good accuracy. Observations in \ensuremath{y_{\rm P1}}\ of the fainter white dwarfs 
were curtailed at 100 sec duration, so their uncertainties are relatively large. In addition, Medium Deep Field 09 (which overlaps SDSS Stripe 82) 
was observed a dozen times in each of \ensuremath{g_{\rm P1}}\, \ensuremath{r_{\rm P1}}\, \ensuremath{i_{\rm P1}}\, \ensuremath{z_{\rm P1}}\, and \ensuremath{y_{\rm P1}}\, offering the opportunity to tie the spectrophometric data to a well-observed 
Pan-STARRS1 field. All stars were placed on OTA 34 and cell 33, so their integration was on the same silicon and used the same 
amplifier for read out (gain measured to be 0.97 e$^-$/ADU).
The observations were bias subtracted and flatfielded as part of the normal IPP processing, and the IPP fluxes (instrumental magnitudes) were 
then available for comparison with tabulated SEDs. The IPP performs an aperture correction and reports fluxes within a radius of 
25 pixels (13 arcsec diameter).  Observations of Polaris on MJD 55744 with the spectroscopic sky probe had a PWV indistinguishable from the 
long term mean of 0.65 cm.

\section{Putting it all Together: Closing the Photometric Loop and Assessing Consistency}

As described in more detail in \cite{JT12}, we compared the observed \ensuremath{g_{\rm P1}}\, \ensuremath{r_{\rm P1}}\, \ensuremath{i_{\rm P1}}\, \ensuremath{z_{\rm P1}}\ and \ensuremath{y_{\rm P1}}\ fluxes for the 
HST spectrophotometric standards with the predictions obtained from using equation (1), using our 
in-dome determination of the instrumental response, the observationally adjusted MODTRAN model for the atmosphere, 
and the HST Calspec data for the source SEDs. We find that we need to apply a gently varying ``tweak'' to the system response function, 
with an rms of around 3-4\%, in order to obtain consistency between the synthetic photometry using CalSpec SEDs and the on-sky observations. 

\section{Conclusions}

The results presented here comprise, to our knowledge, the first instance of a NIST-calibrated telescope response function and a 
MODTRAN atmospheric model generating synthetic photometry that is then compared with on-sky measurements of spectrophotometric standards. 
We obtain agreement at the 5\% level, except for the \ensuremath{y_{\rm P1}}\ band where the discrepancy is 10\%. We consider this to be encouraging, 
since it's our first attempt to ``close'' this photometric loop.  Our objectives for future work include:

\begin{enumerate}

\item{} Obtaining more precise on-site determination of atmospheric aerosols. 

\item{} Improving the in-dome calibration technique, and our corrections for optical multipath effects. 

\item{} Identifying the origin of the ``tweak'' we need to apply to obtain overall consistency between the 
NIST-based calibration and the expectations from stellar SEDs. 

\end{enumerate}

\acknowledgements 

We are grateful to the US National Science Foundation, under grants AST-0443378, 
AST-0507475 and AST-1009749, and AST-0551161 (awarded for
LSST development). We are also 
grateful to the LSST Corporation, Harvard University, and the US Department of 
Energy Office of Science
for their support of the continuing development of these
techniques under grant DE-SC0007881 and NIST under grant 70NANB8H8007. We thank the following colleagues
for their contributions and collaboration on this overall effort: 
Justin Albert (U Victoria),
Tim Axelrod (U Arizona)
Ralph Bohlin (STSCI),
Steve Brown (NIST),
Yorke Brown (Dartmouth),
David Burke (SLAC),
Ken Chambers (UH),
Claire Cramer (NIST),
Susana Deustua (STSCI), 
Peter Doherty (Harvard),
Maxwell Fagin (Dartmouth),
Will High (Univ. Chicago),
Keith Lykke (NIST),
John McGraw (UNM),
Gautham Narayan (Harvard),
Abi Saha (NOAO),
Isaac Shivvers (Berkeley),
Chris Smith, CTIO),
Amali Vaz (Harvard), 
John Woodward (NIST),
and Peter Zimmer (UNM).

We also thank the conference organizers for their hard work in arranging a most interesting and informative meeting.

\bibliography{stubbs_c}

\section*{Discussion}

Q: (S. Deustua) What about flatfields? How reliable is your flat/illumination system in that a source placed
on any pixel gives the same result? You have a very wide (3 deg) field. 

\noindent
Q: (N. Regnault) It is difficult to apply a photometric calibration obtained from a flat-field-like illumination to point source photometry. 
Do you have specific plans to bridge this gap? 

\noindent
A: We agree that it's difficult to distinguish the focusing light paths from the stray and scattered light that lands on the focal plane, 
when the illumination comes from a diffuse flat field screen. As described above, we have used a collimated beam to attempt to 
disentangle these illumination paths, and to determine the response function at one location on the CCD array. The task of tying together the 
system sensitivity across the array is a distinct but related problem. A number of groups are working to develop various calibration devices that control the 
phase space distribution of the rays (in angle and position) that are sent into the telescope pupil from a monitored calibration source. 
We think that making an ``illumination correction'' is an essential ingredient in reducing and assessing sources of systematic error. 
This is best obtained on the sky, either by rastering stable sources across the focal plane, or through the ``ubercal'' procedure that uses
the multiple observations per field from a wide-field multi-epoch survey.  As a community we are still developing an overall 
methodology for obtaining the best possible accuracy and precision, and this meeting has been very useful in helping move this forward.

\end{document}